\documentstyle{article}
\onecolumn
\def\dj{d\kern-.35em\raise1.25ex\vbox{\hrule width .3em height
.03em}\kern.05em}
\def\Dj{D\kern-.70em\raise0.75ex\vbox{\hrule width .3em height
.03em}\kern.40em}

\def\^{\hat}

\def\ra{\rightarrow}

\def\w{\omega}

\begin{document}

\abovedisplayskip       1ex
\belowdisplayskip \abovedisplayskip
\abovedisplayshortskip  0ex
\belowdisplayshortskip  1ex

\partopsep=0in
\itemsep  =0in
\topsep   =0in
\parsep   =0in

\begin{center}
{\Large QUANTUM PRINCIPAL BUNDLES}\\
by Mi\'co \Dj ur\dj evi\'c
\end{center}
\begin{center}
{\em Faculty of Physics, P.O.BOX 550, Studentski Trg 12, Beograd,
Serbia, Yugoslavia}
\end{center}

       {\em Abstract}: A noncommutative-geometric generalization of
the
theory of principal bundles is sketched. A  differential  calculus
over corresponding quantum  principal  bundles  is  analysed.  The
formalism of connections is presented. In particular, operators of
covariant derivative and horizontal projection are  described  and
analysed. Quantum counterparts for the Bianchi  identity  and  the
Weil's homomorphism are found. Illustrative examples are considered.

     I INTRODUCTION

     The purpose of this letter is to present the  most  important
elements of a quantum generalization of the  theory  of  principal
bundles, in which quantum groups play the role of structure groups,
and quantum spaces the role of base manifolds. All  considerations
are performed within  the  conceptual  scheme  of  non-commutative
differential geometry [C1,2]. A detailed exposition of the theory  is
given in papers [D1,2].

     The paper is organized as follows. Section II begins  with  a
definition of quantum principal bundles. Then,  questions  related
to differential calculus are discussed. Section III is devoted  to
the formalism of connections. In Section IV  a  generalization  of
the Weil's theory of characteristic classes is  sketched. Finally,
in Section  V some examples  of  quantum  principal bundles are
considered.

     Before passing to quantum principal bundles we shall fix  the
notation, and introduce the relevant quantum group entities. Here,
we shall deal with compact matrix quantum groups [W2].  Let  $G$
be such a group. The algebra of 'polynomial functions' on  $G$  will
be denoted by ${\cal A}$. The group structure on $G$  is  determined  by  the
comultiplication $\phi:{\cal A}\ra{\cal A}\otimes{\cal A}$,
the  counit $e:{\cal A}\ra C$,  and  the antipode
$k:{\cal A}\ra{\cal A}$. The result of the action of an n-fold
comultiplication  on
elements $a\in {\cal A}$ will be symbolically written as
$a^{(1)}\otimes ...\otimes a^{(n)}$.  We
shall denote by $ad:{\cal A}\ra {\cal A}\otimes {\cal A}$
the  adjoint  action  of  $G$  on  itself.
Explicitly, this action is given by
                     \[ad(a)=a^{(2)}\otimes k(a^{(1)})a^{(3)}.\]

     Let $(\Gamma ,d)$ be a first-order differential calculus [W3] over $G$,
and let
 \[\Gamma ^{\wedge}={\sum_{k\geq 0}}^\oplus \Gamma ^{\wedge k}\]
     be the universal differential envelope of $ (\Gamma ,d)$ [D1]  (with
$\Gamma ^{\wedge 0} = {\cal A}$
and $\Gamma ^{\wedge 1}=\Gamma$ ). For  each $k\geq 0$  let
$p_k :\Gamma ^{\wedge} \ra \Gamma ^{\wedge k}$
be  the  corresponding projection. Further, let
\[\Gamma ^{\otimes}={\sum_{k\geq 0}}^\oplus \Gamma ^{\otimes k}\]
be the tensor bundle algebra over $\Gamma$ ($\Gamma ^{\otimes k}=
\Gamma \otimes_{\cal A}...\otimes_{\cal A} \Gamma$  (k-times)  and
$\Gamma ^{\otimes 0}={\cal A}$). Let us assume that $(\Gamma ,d)$
is left-covariant. We shall denote
by $\Gamma _{inv}$    the space of left-invariant  elements of $\Gamma$
while ${\cal R} \subseteq ker(e) $
will be the right ${\cal A}$-ideal which canonically, in the sense of  [W3]
corresponds to $(\Gamma ,d)$. The map $\pi :{\cal A}\ra \Gamma_{inv}$
given by
                       \[\pi (a)=k(a^{(1)})da^{(2)}\]
is surjective, and $ker(\pi )=C1\oplus {\cal R}$. Because of this,
there  exists  a
natural isomorphism
                     \[\Gamma_{inv} =ker(e)/{\cal R}.\]

The above isomorphism induces a right ${\cal A}$-module structure on
$\Gamma_{inv}$ ,
which will be denoted by $\circ$ . Explicitly,
                      \[\pi (a) \circ b=\pi (ab),\]
for each $a\in ker(e)$ and $b \in {\cal A}$. The tensor product of
$k$ copies of  $\Gamma _{inv}$
will be denoted by $\Gamma _{inv} ^{\otimes k}$. The tensor  algebra
over  $\Gamma_{inv}$     will  be denoted by $\Gamma_{inv} ^{\otimes}$.
It  is
naturally  isomorphic  to  the  space  of left-invariant elements
of $\Gamma ^{\otimes}$. The  subalgebra  of  left-invariant elements
of $\Gamma ^{\wedge}$  will be denoted by $\Gamma ^{\wedge} _{inv}$.
We have
\[\Gamma ^{\wedge} _{inv}={\sum_{k\geq 0}}^\oplus \Gamma_{inv}^{\wedge k}\]
where $\Gamma_{inv}^{\wedge k}$    consists of left-invariant
elements of $k$-th degrees. For
each $k\geq 0$ let $\Pi _{inv}^k:\Gamma^{\wedge k}\ra \Gamma^{\wedge k}_{inv}$
be the projection onto  left-invariant
elements  (characterized  by $\Pi _{inv}^k(a\vartheta)=e(a)\vartheta $
for  each
$a\in {\cal A}$   and $\vartheta\in \Gamma^{\wedge k}_{inv}$), and let
$p_{inv}^k:\Gamma^{\wedge}_{inv} \ra \Gamma^{\wedge k}_{inv}$
be the  standard  projection  map.
The following natural isomorphism holds
\[\Gamma^{\wedge}_{inv}=\Gamma^{\otimes}_{inv}/I_{inv}^{\wedge}.\]
\filbreak

     Here $I_{inv}^{\wedge}\subseteq \Gamma^{\otimes}_{inv}$
is the ideal generated by elements of the form
$\pi (a^{(1)})\otimes \pi (a^{(2)})$ where $a\in {\cal R}$.
The right ${\cal A}$-module structure $\circ$ can be
uniquely extended from $\Gamma_{inv}$ to $\Gamma_{inv}^{\wedge,\otimes}$
such that
                         \[1\circ a=e(a)1\]
\[(\vartheta \eta )\circ a=(\vartheta \circ a^{(1)})(\eta \circ a^{(2)})\]
for each $\vartheta ,\eta \in \Gamma_{inv}^{\wedge ,\otimes}$ and
$a \in {\cal A}$.

     Let us now assume that the calculus $(\Gamma ,d)$ is bicovariant, and
let $\widetilde{ad}:\Gamma_{inv} \ra \Gamma_{inv} \otimes {\cal A}$ be
the adjoint
action of $G$ on $\Gamma_{inv}$ (coinciding
with the restriction of the right action of $G$ on $\Gamma_{inv}$ ). We have
\[\widetilde{ad}\pi =(\pi \otimes id)ad.\]

In the following, we shall denote by
$\widetilde{ad}^{\otimes},\widetilde{ad}^{\otimes k},
\widetilde{ad}^{\wedge},\widetilde{ad}^{\wedge k}$ the
adjoint actions of  $G$ on the corresponding spaces.

     The  map $\phi :{\cal A} \ra {\cal A} \otimes {\cal A}$
admits  the  unique   extension   to   the homomorphism
$\hat \phi :\Gamma ^{\wedge} \ra \Gamma^{\wedge}
{\hat \otimes} \Gamma^{\wedge}$
of (graded) differential algebras.

     II QUANTUM PRINCIPAL BUNDLES AND THE CORRESPONDING
         DIFFERENTIAL CALCULUS

     The aim of this section is  to introduce quantum  principal
bundles, and to describe differential calculus over them.

     Let $M$ be a quantum space, represented by a *-algebra ${\cal V}$.  The
elements of ${\cal V}$ play the role of appropriate 'functions' on $M$.

     DEFINITION 2.1. {\em A quantum  principal  $G$-bundle}
over  $M$  is a
triplet of the form $P=({\cal B},i,F)$ where ${\cal B}$ is a *-algebra,
while $F: {\cal B} \ra {\cal B} \otimes {\cal A}$
and $i:{\cal V} \ra {\cal B}$ are unital *-homomorphisms such that
     \begin{itemize}
    \item[(i)] The following identities hold
\[id=(id\otimes e)F\]
                      \[(id\otimes \phi)F=(F\otimes id)F.\]
    \item[(ii)] The map $i:{\cal V} \ra {\cal B}$ is injective and
$b\in i({\cal V})$ iff $F(b)=b \otimes 1$,
for each $b\in{\cal B}$.
   \item[(iii)] A linear map
$X:{\cal B}\otimes {\cal B} \ra {\cal B} \otimes {\cal A}$ defined by
                        \[X(a\otimes b)=aF(b)\]
is surjective.
\end{itemize}

    The map $F$ plays the role of the dualized right action   of $G$
on $P$. Condition (i) justifies this interpretation. The map
$i:{\cal V}\ra {\cal B} $
can be interpreted as the dualized projection on $M$. Condition (ii)
says that $M$ can be identified with the corresponding 'orbit space'
of $P$. Finally, condition (iii) is an effective quantum counterpart
of the classical requirement that $G$ acts freely on $P$.

    Let $P=({\cal B},i,F)$ be a quantum principal $G$-bundle over $M$.

    We are going to construct a graded  differential  algebra
representing verticalized differential forms on $P$. Let us fix a
bicovariant first-order differential *-calculus $(\Gamma ,d)$  over  $G$.
The *-involution naturally extends from $\Gamma$ to
$\Gamma^{\wedge ,\otimes}$ (such  that
$(\vartheta \eta)^* = (-1)^{deg(\vartheta )deg(\eta )} \eta^* \vartheta^*$
for each $\vartheta ,\eta \in \Gamma^{\wedge ,\otimes}$). Algebras
$\Gamma^{\wedge ,\otimes}_{inv} \subseteq \Gamma^{\wedge ,\otimes}$
are *-invariant.

    Let us consider the (graded) vector space
$ver(P)={\cal B}\otimes \Gamma^{\wedge}_{inv}$.

    LEMMA 2.1. {\em The formulas }

\[(q\otimes \vartheta )(b \otimes \eta)=\sum_{k} qb_k \otimes
(\vartheta \circ c_k )\eta \]
\[(b\otimes \eta)^* = \sum_{k} b_k^* \otimes (\eta^* \circ c_{k}^*)\]
\[d_v (b\otimes \eta )= b\otimes d\eta + \sum_{k} b_k \otimes \pi (c_k)\eta\]
{\em where} $F(b)=\sum_{k} b_k \otimes c_k$
{\em determine the structure of a graded differential *-algebra on} $ver(P)$.
{\em As a differential algebra,} $ver(P)$ {\em is generated by}
${\cal B}=ver^0 (P)$.$\Box$
\filbreak

     We shall assume that a differential calculus over the bundle $P$
is specified by a graded differential *-algebra $\Omega (P)$ such that
    \begin{itemize}
    \item[(i)] The differential algebra $\Omega (P)$ is
generated by ${\cal B}=\Omega^0 (P)$.
    \item[(ii)] The map $F:{\cal B} \ra {\cal B} \otimes {\cal A}$
is extendable to a homomorphism
\[\hat F :\Omega (P) \ra \Omega (P) \hat {\otimes} \Gamma^{\wedge}\]
of (graded) differential algebras.
     \end{itemize}

     The map $\hat F$ is uniquely determined by the above conditions.
We have
\[(\hat F\otimes id)\hat F=(id\otimes {\hat \phi})\hat F.\]
The formula
\[F^{\wedge} =(id\otimes p_0)\hat F\]
defines the action $F^{\wedge}:\Omega (P) \ra \Omega (P) \otimes {\cal A}$ of
G on differential forms (extending the action $F$). The map $F^\wedge$  is a
*-homomorphism
and
                          \[(id\otimes e)F^{\wedge} =id\]
\[(F^{\wedge}\otimes id)F^{\wedge} =(id\otimes\phi)F^{\wedge}\]
\[F^{\wedge}d=(d\otimes id)F^{\wedge}.\]

     Let us  construct  a  quantum  analog  of  the  verticalising
homomorphism. For each $w\in \Omega^k (P)$ the element
$(id\otimes \Pi_{inv} ^k p_k )\hat F (w)$ belongs
to ${\cal B}\otimes \Gamma_{inv}^{\wedge k} =ver^k (P)$.
Hence, the formula
$\pi_v (w)=(id\otimes \Pi_{inv}^k p_k )\hat F (w)$
defines a linear grade-preserving map
\[\pi_v :\Omega (P) \ra ver(P).\]

     LEMMA 2.2. {\em The map} $\pi_v$  {\em is an epimorphism  of graded
differential *- algebras.}$\Box$

     Now, horizontal forms will be defined. Intuitively  speaking,
they can be characterized as forms possessing trivial differential
properties along vertical fibers.

    DEFINITION 2.2. The elements of the graded *-subalgebra
\[hor(P)=\hat F^{-1}[\Omega(P)\otimes {\cal A}]\]
of $\Omega(P)$ are called {\em horizontal forms}.

     Horizontal forms $w$ satisfying $F^{\wedge} (w)=w\otimes 1$
are  interpretable as differential forms on $M$. They
constitute  a  graded  differential *-subalgebra
$\Omega (M) \subseteq \Omega (P)$, with $\Omega^0 (M)=i({\cal V})$.

     III THE FORMALISM OF CONNECTIONS

     Before introducing connections in the game, we shall define
(pseudo)tensorial forms.

     Let $\psi (P)$ be the space of linear maps
$f:\Gamma_{inv} \ra \Omega (P)$ satisfying
                \[F^{\wedge}f=(f\otimes id)\widetilde{ad}.\]

        This space is naturally graded. The  elements  of $\psi^k (P)$
are imaginable as {\em pseudotensorial $k$-forms} on $P$, with values from
the 'lie algebra' of $G$.  Further, $\psi (P)$  is  closed with respect
to compositions with $d:\Omega (P) \ra \Omega (P)$. Let
\[\tau (P)=\{f\in \psi (P):f(\Gamma_{inv})\subseteq hor(P)\}\]
be the graded subspace of $\psi (P)$ consisting of {\em tensorial forms}.

        The formula
                        \[f^*(\vartheta)=f(\vartheta^*)^*  \]
determines a *-involution on $\psi (P)$ (and $\tau (P)$).

     DEFINITION 3.1. {\em A connection on} $P$ is a linear map
$\omega :\Gamma_{inv}\ra \Omega^1 (P)$ such that
\[\hat F \omega (\vartheta )=(\omega \otimes id)\widetilde{ad}(\vartheta)
+1\otimes \vartheta\]
\[\omega (\vartheta^*)=\omega(\vartheta)^*,\]
for each $\vartheta \in \Gamma_{inv}$.
\filbreak

    Connections can be equivalently defined as hermitian pseudotensorial
one-forms $\omega$ satisfying
\[\pi_v \omega (\vartheta)=1\otimes \vartheta\]
for each $\vartheta \in \Gamma_{inv}$.
\filbreak

    LEMMA 3.1. {\em The bundle} $P$ {\em admits at least one connection.}$\Box$

    Let $con(P)$ be the set of all connections on $P$. This is
is a real affine  subspace  of $\psi^1 (P)$.  The  corresponding  vector
space consists of hermitian tensorial 1-forms.

     Let us fix  a  linear  map
$\delta :\Gamma_{inv} \ra \Gamma_{inv}^{\otimes 2}$     with  the  following
properties
\[\delta *=-(*\otimes *)\delta\]
\[(\delta \otimes id)\widetilde{ad}=\widetilde{ad}^{\otimes 2}\delta\]
\[d\vartheta=\sum_{k}\vartheta_k^1\vartheta_k^2\]
where $\vartheta \in \Gamma_{inv}$ and
$\delta (\vartheta)=\sum_{k} \vartheta_k^1 \otimes \vartheta_k^2$.

    For given linear maps $\varphi,\eta :\Gamma_{inv} \ra \Omega (P)$
let us define  new  linear  maps
$<\varphi,\eta>,[\varphi,\eta]:\Gamma_{inv}\ra\Omega(P)$ by
\[<\varphi,\eta>=m_{\Omega}(\varphi\otimes\eta)\delta\]
\[[\varphi,\eta]=m_{\Omega}(\varphi\otimes\eta)c^T\]
where
$c^T =(id\otimes\pi)\widetilde{ad}:\Gamma_{inv} \ra \Gamma_{inv}^{\otimes 2}$
and $m_{\Omega}:\Omega (P)\otimes \Omega (P)\ra\Omega (P)$
are the 'transposed  commutator'  [W3] and the multiplication map.

    If $\varphi\in\psi^i (P)$  and $\eta\in\psi^j (P)$   then
$<\varphi,\eta>,[\varphi,\eta]\in\psi^{i+j}(P)$.

    For each $\omega\in con(P)$ let us consider a map
\[R_{\omega} =d\omega-<\omega,\omega>.\]

    LEMMA 3.2.{\em We have}
 \[\hat F R_{\omega} =(R_{\omega}\otimes id)\widetilde{ad}\]
 \[R_{\omega}^* =R_{\omega}.\]
  {\em In other words, $R_\omega$ is a tensorial hermitian 2-form}.$\Box$

    DEFINITION 3.2. The map $R_\omega$  is called {\em the curvature of}
    $\omega$.

    It is worth noticing that $R_\omega$  depends on the choice of $\delta$.
This dependence disappears if $\omega$ satisfies the following multiplicativity
property.

     DEFINITION 3.3. A connection $\omega$ is called {\em multiplicative} iff
\[\omega\pi(a^{(1)})\omega\pi(a^{(2)})=0\]
for each $a\in{\cal R}$.

    If $\omega$ is multiplicative then it can be  uniquely  extended,  by
multiplicativity, to a unital (*-) homomorphism
$\omega^{\wedge}:\Gamma_{inv}^{\wedge}\ra \Omega (P)$.
Another interesting class of connections consists  of those having
the  following regularity property.

    DEFINITION 3.4. A connection $\omega$ is called {\em regular} iff
\[\omega (\vartheta)\varphi=
(-1)^{deg(\varphi)}\sum_{k}\varphi_k\omega(\vartheta\circ c_k)\]
for each $\vartheta\in\Gamma_{inv}$ and $\varphi\in hor(P)$, where
$F^{\wedge}(\varphi)=\sum_{k} \varphi_k\otimes c_k$.

    Regular connections (if exist) form an affine subspace $conr(P) $
of $con(P)$. The corresponding vector space  consists  of
forms $f=f^*\in\tau^1 (P)$ satisfying
\[f(\vartheta)\varphi=
(-1)^{deg(\varphi)}\sum_{k}\varphi_k f(\vartheta\circ c_k)\]
for each $\vartheta\in\Gamma_{inv}$ and $\varphi\in hor(P)$.

     Let $\sigma :\Gamma_{inv}^{\otimes 2}\ra\Gamma_{inv}^{\otimes 2}$
be the canonical flip-over operator [W3].
\filbreak

     LEMMA 3.3. {\em If} $\omega\in conr(P)$ {\em then }
\[m_{\Omega}(\omega\otimes \varphi)=
(-1)^k m_{\Omega}(\varphi\otimes\omega)\sigma\]
{\em for each} $\varphi \in \tau^k (P)$.$\Box$

     Now, we are going to  introduce  the  operator  of  covariant
derivative. This operator will be first defined on a restricted
domain consisting of  horizontal  forms. After  introducing  the
operator of horizontal projection, the  domain  of
covariant derivative will be extended to the whole algebra $\Omega (P)$.

     For each $\omega\in con(P)$ and $\varphi\in hor(P)$
let us define a new form
\[D_{\omega} (\varphi)=d\varphi
-(-1)^{deg(\varphi)}\sum_{k} \varphi_k \omega\pi(c_k), \]
where $F^{\wedge}(\varphi)=\sum_{k} \varphi_k\otimes c_k$.

     The form $D_{\omega}(\varphi)$ is horizontal, too.

     DEFINITION 3.5. A linear map $D_{\omega} :hor(P)\ra hor(P)$
is called {\em the covariant derivative} associated to $\omega$.

     PROPOSITION 3.4. \begin{itemize}
     \item[(i)] {\em The map $D_{\omega}$ intertwines the action
$(F^\wedge\mid hor(P)):hor(P)\ra hor(P)\otimes{\cal A}$ with itself.}
     \item[(ii)] {\em If} $\omega$ {\em is multiplicative then}
\[D_{\omega}^2 (\varphi)=-\sum_k \varphi_k R_{\omega}\pi (c_k),\]
{\em for each} $\varphi\in hor(P)$.
    \item[(iii)] {\em If $\omega$ is regular then}
\[D_{\omega}(\varphi\psi)=D_{\omega}(\varphi)\psi +(-1)^{deg(\varphi)}
\varphi D_{\omega} (\psi) \]
\[D_{\omega} (\varphi^*)=D_{\omega}(\varphi)^*\]
{\em for each} $\varphi ,\psi \in hor(P)$.
     \item[(iv)] {\em If $\varphi\in\Omega (M)$ then
$D_{\omega} (\varphi)=d\varphi$.}$\Box$
\end{itemize}

     The space $\tau (P)$ is  closed  under  taking
compositions with $D_{\omega}$. This fact enables us to define the
action of covariant derivative in the space of tensorial forms.

     LEMMA 3.5. {\em We have
\[D_{\omega} (\varphi)=d\varphi -(-1)^{deg(\varphi)}[\varphi ,\omega]\]
for each} $\varphi\in\tau (P)$.$\Box$

     Let us consider a linear map $q_{\omega} :\psi (P)\ra \psi (P)$
defined by
\[q_{\omega} (\varphi)=<\omega,\varphi>-
(-1)^{deg(\varphi)}<\varphi,\omega>
-(-1)^{deg(\varphi)}[\varphi,\omega].\]
We have then
\[\hat F q_{\omega} (\varphi)=
(q_{\omega} \otimes id)F^{\wedge}(\varphi) \]
for each $\varphi\in\tau (P)$. In particular,$ q_{\omega}\tau (P)\subseteq
\tau (P)$. Moreover, if $\omega\in conr(P)$ then $(q_{\omega}\mid\tau (P))=0$.

     The following lemma gives the  quantum  counterpart  for  the
classical Bianchi identity.

     LEMMA 3.6. {\em We have}
\[(D_{\omega} -q_{\omega})(R_{\omega})=
<\omega,<\omega,\omega>>-<<\omega,\omega>,\omega> \]
{\em for each} $\omega \in con(P)$.$\Box$

     If the connection $\w$ is multiplicative, then  the  right hand
side of the above equality vanishes. On the other hand,  if $\w$  is
regular then the second summand in the left hand side vanishes. It
is worth noticing that regular  connections  are  not  necessarily
multiplicative. However, there  exists  a  common  obstruction  to
multiplicativity for all regular connections,  so  that  they  are
multiplicative, or not, at the same time.
\filbreak

     In general, the lack of multiplicativity of the connection $\omega$
is measured by a map $r_{\omega} :{\cal R}\ra \Omega^2 (P)$ given by
$r_{\omega}(a)=\omega\pi(a^{(1)})\omega\pi(a^{(2)})$.

     LEMMA 3.7. \begin{itemize}
     \item[(i)] {\em The following identities hold}
\[r_{\omega} (k(a)^*)=-r_{\omega}(a)^*\]
\[\pi_v r_{\omega}=0 \]
\[\hat F r_{\omega}(a)=(r_{\omega}\otimes id)ad(a). \]
{\em In particular, $r_{\omega} (a)$ is horizontal for each}
$a\in {\cal R}$.
   \item[(ii)] {\em The map $\omega \mapsto r_{\omega}$
is constant on cosets from the space} $con(P)/conr(P)$. {\em If
$\omega \in conr(P)$ then }
\[r_{\omega}(a)\varphi =\sum_{k}\varphi_k r_{\omega}(ac_k)\]
{\em for each $a\in {\cal R}$ and $\varphi\in hor(P)$,
where $F^{\wedge}(\varphi)=\sum_{k} \varphi_k\otimes c_k$. Further,}
\[dr_{\omega}(a)=<\omega ,\omega>\pi(a^{(1)})\omega\pi(a^{(2)})
-\omega\pi(a^{(1)})<\omega,\omega>\pi(a^{(2)}).\Box\]
\end{itemize}

    Let us assume that $P$ admits regular connections, and let $ J(P) $
be the ideal in $\Omega (P)$ generated by the  space
$r_{\omega}({\cal R})$,  for some  $\omega\in conr(P)$.  The previous lemma
implies
\[J(P)^* =J(P)\]
\[\hat F J(P)\subseteq J(P)\otimes \Gamma^{\wedge}\]
\[\pi_v J(P)=\{0\}\]
\[dJ(P)\subseteq J(P).\]

    Consequently, it is possible to project the whole formalism on the
factoralgebra $\Omega (P)/J(P)$.
In  the  framework  of  this  projected  calculus  regular
connections become  multiplicative.

    The last  topic  in  this  section  is  the  construction  and
the analysis  of  horizontal  projection  operators.  Let us fix a
splitting of the form
\[\Gamma_{inv}^{\otimes}=
\Gamma_{inv}^{\wedge}\oplus I_{inv}^{\wedge}\]
in which $\Gamma_{inv}^{\wedge}$ is realized as a
complement of the space  $I_{inv}^{\wedge}$,      with
the help of a grade-preserving hermitian section
$\iota :\Gamma_{inv}^{\wedge} \ra \Gamma_{inv}^{\otimes}$,
intertwining  the
adjoint actions. Further, let us assume that
$\delta(\vartheta)=\iota d(\vartheta)$. Finally,  let
us consider a linear map
$m_{\omega} :hor(P)\otimes\Gamma_{inv}^{\wedge} \ra \Omega (P)$ given by
\[m_{\omega}(\varphi \otimes \vartheta)=
\varphi \omega^{\otimes}\iota (\vartheta).\]
Here, $\omega^{\otimes} :\Gamma_{inv}^{\otimes} \ra \Omega (P)$
is the unital multiplicative extension of $\omega$. It is worth noticing
that in the special case of {\em multiplicative} connections the map
$m_\omega$ is $\iota$-independent, because $\omega^{\otimes}\iota=
\omega^\wedge$ in this case.

    PROPOSITION 3.8. \begin{itemize}
    \item[(i)] {\em The map $m_{\omega}$ is bijective. It intertwines the
product of actions $(F^{\wedge}\mid hor(P))$ and
$\widetilde{ad}^{\wedge}$,  with the action} $F^{\wedge}$.
   \item[(ii)] {\em If $\omega$ is regular and if $J(P)=\{0\}$
then $m_{\omega}$
is an isomorphism of  *-algebras.
Here, it is assumed that $hor(P)\otimes\Gamma_{inv}^{\wedge}$
is endowed  with  a (graded) *-algebra structure specified by
\[(\psi\otimes\eta)(\varphi\otimes\vartheta)=
\sum_{k} (-1)^{deg(\varphi)deg(\eta)}\psi\varphi_k \otimes (\eta\circ c_k)
\vartheta\]
\[(\varphi\otimes\vartheta)^*=\sum_{k} \varphi_k^* \otimes
\vartheta^* \circ c_k^*,\]
where} $F^{\wedge}(\varphi)=\sum_{k} \varphi_k\otimes c_k$.$\Box$
\end{itemize}

    The horizontal projection operator
$h_{\omega} :\Omega (P) \ra hor(P)$  can  be now
defined as follows
\[h_{\omega} =(id\otimes p_{inv}^0)m_{\omega}^{-1}\]
Clearly, $h_{\omega}$  projects $\Omega (P)$ onto $hor(P)$.
\filbreak

    With the help of $h_{\omega}$, the domain of the
covariant derivative can be  extended
to the whole algebra $\Omega (P)$. Indeed, the map
$D_{\omega} :\Omega (P) \ra hor(P)$ given by
\[D_{\omega}=h_{\omega}d\]
extends the previously defined covariant derivative.

    PROPOSITION 3.9. \begin{itemize}
    \item[(i)] {\em The maps $h_{\omega},D_{\omega}$ intertwine
the actions $F^{\wedge}$ and} $(F^{\wedge}\mid hor(P))$.
    \item[(ii)] {\em If $\omega\in conr(P)$ then $h_{\omega}$
is a *-homomorphism, and
\[D_{\omega}(ww')=D_{\omega}(w)h_{\omega}(w')+
h_{\omega}(w)D_{\omega}(w') \]
\[D_{\omega}(w^*)=D_{\omega}(w)^*\]
for each} $w,w'\in \Omega (P)$.$\Box$
\end{itemize}

    Compositions of pseudotensorial forms with the covariant derivative
are  tensorial. Consequently, it is possible to define the
covariant  derivative $D_{\omega} :\psi (P) \ra \tau (P)$. The following
lemma gives an equivalent, more geometrical,
description  of the curvature.

    LEMMA 3.10. {\em We have
\[R_{\omega} =D_{\omega}(\omega)\]
for each} $\omega\in con(P)$.$\Box$

    IV CHARACTERISTIC CLASSES

    In  this  section  we  shall sketch a   quantum
generalization of the Weil's  theory  of  characteristic  classes.
We shall assume  that
the bundle $P$ admits regular connections, and that $J(P)=\{0\}$.
    For each $k\geq 0$ let $Inv^k \subseteq \Gamma_{inv}^{\otimes k}$
be the subspace  of $ad$-invariant elements, and let
$Inv$ be the direct sum of all these  spaces. Clearly, $Inv$
is a unital *-subalgebra of the tensor algebra
$\Gamma_{inv}^{\otimes k}$. Let $H(M)$ be the graded *-algebra of
cohomology classes associated to $\Omega (M)$.

    Let us consider a connection $\omega$. There exists the unique unital
homomorphism $R_{\omega}^{\otimes}:\Gamma_{inv}^{\otimes}\ra \Omega (P)$
extending the curvature $R_{\omega}$. The  map $R_{\omega}^{\otimes}$
is *-preserving, and intertwines $\widetilde{ad}^{\otimes}$ and $F^{\wedge}$.

    PROPOSITION 4.1. \begin{itemize}
    \item[(i)] {\em If $\vartheta\in Inv^k$ then
$R_{\omega}^{\otimes}(\vartheta)\in \Omega^{2k} (M)$. }
    \item[(ii)] {\em If $\omega \in conr(P)$ then
$dR_{\omega}^{\otimes}(\vartheta)=0$ for each $\vartheta \in Inv$.}
   \item[(iii)] {\em The cohomological class of
$R_{\omega}^{\otimes}(\vartheta)$ in $\Omega (M)$ is
independent
on the choice of a regular connection $\omega$, for each
$\vartheta \in Inv$.}
    \item[(iv)] {\em The map $W:Inv\ra H(M)$ given  by
$W(\vartheta)=[R_{\omega}^{\otimes}(\vartheta)] $ is
a  unital *-homomorphism.}$\Box$
\end{itemize}

    The homomorphism $W$ plays the role of the  Weil's  homomorphism
in classical differential geometry [KN]. In fact, in  classical
geometry  the  domain  of  the  Weil's
homomorphism is restricted on the algebra of  symmetric  invariant
elements of the corresponding  tensor  algebra.  However,  besides
simplifying the domain of $W$, such a restriction gives nothing new:
the image of the Weil's homomorphism will be the same.

    A similar situation holds  in  the  noncommutative  case.  Let
$Sym(\sigma)$ be the *-algebra obtained from $\Gamma_{inv}^{\otimes}$
by  factorising  through
the ideal $I(\sigma)$  generated  by
$Im(I-\sigma )\subseteq \Gamma_{inv}^{\otimes 2}$.  The  algebra
$Sym(\sigma) $
plays the role of polynoms over the 'lie  algebra'  of  $G$.  The  adjoint
action $\widetilde{ad}^{\otimes}$  is naturally projectable on
$Sym(\sigma)$.  Let  $Inv(\sigma)\subseteq Sym(\sigma) $
be the  subalgebra  of  elements  invariant  under  the  projected
action (playing the role of invariant polynomials). Clearly,
$Inv(\sigma)=Inv/(I(\sigma)\cap Inv)$.

    PROPOSITION 4.2. {\em If $\omega \in conr(P)$ then
\[R_{\omega}^{\otimes}\sigma(\vartheta)=R_{\omega}^{\otimes}(\vartheta)\]
for each} $\vartheta\in\Gamma_{inv}^{\otimes 2}$.$\Box$

    The above statement implies that $W$ and $R_{\omega}^{\otimes}$
can  be  factorised through the ideal $I(\sigma)$.
\filbreak

    V EXAMPLES AND REMARKS

    (A) All quantum phenomena characteristic for
the presented theory of quantum principal bundles already
figure in a special
version of this theory dealing with bundles over
classical smooth manifolds. The
theory of principal bundles of this kind is developed in [D1].

The main structural  result  is  that  $G$-bundles  $P$ over a classical
manifold $ M$ are  in  a  natural
correspondence with classical bundles $P_{cl}$ over the same  manifold,
with the structure group $G_{cl}$ consisting of {\em classical points}
of  $G$.
More precisely, the elements of $G_{cl}$   are  *-characters
$g:{\cal A}\ra C$. The
product and the inverse in $G_{cl}$ are given by
\[gg'=(g\otimes g')\phi\]
\[g^{-1}=gk,\]
while the counit $e:{\cal A}\ra C$ is the neutral element. The
correspondence $P\leftrightarrow P_{cl}$ can  be  roughly
described  as  follows.  The  bundle  $P_{cl}$
consists of classical points of $P$ (*-characters on ${\cal B}$).
Conversely,
if $P_{cl}$ is given then $P$ can be recovered by applying an analog of
the classical construction of extending structure groups.

    In developing a differential calculus on such  semiclassical
bundles $P$ it is natural to assume that all  local  trivializations
of  the  bundle  locally  trivialize  the  calculus,   too.   This
requirement, together with the specification of  the  calculus
$\Gamma^{\wedge}$ over $G$, uniquely fixes the algebra $\Omega (P)$.
However,  the  calculus
$(\Gamma,d)$ can not be chosen  arbitrarily.  It  must  satisfy  specific
consistency  requirements,  interpretable  as  compatibility properties
with certain 'retrivialization maps' of the bundle. Such  differential
calculi  are  called 'admissible' in [D1]. It turns out that a
left-covariant calculus $ (\Gamma,d)$
is admissible iff  $(X\otimes id)ad({\cal R})=\{0\}$, for each
$X\in lie(G_{cl})$. Here, the Lie algebra of $G_{cl}$
is  understood  as  the  space  of  hermitian
functionals $X$ on ${\cal A}$ satisfying  $X(ab)=e(a)X(b)+e(b)X(a)$,
for  each $a,b\in {\cal A}$.

    There exists the minimal admissible  left-covariant  calculus:
it is based on the right-ideal ${\cal R}\subseteq ker(e)$
consisting of elements killed  by  all  operators $(X\otimes id)ad$.
This   calculus   is   also
*-covariant and right-covariant.  If  $G$  is  an  ordinary  compact
matrix group then the minimal admissible calculus  coincides  with
the usual one (based on differential forms). However, small quantum
deformations  of  the  classical   group   structure   may   cause
drastical changes at the level of the minimal  admissible  calculus.
For example [D1], if $G=SU_{\mu}(2)$ [W1] and
$\mu\in (-1,1)\setminus \{0\}$ then the space
$\Gamma_{inv}$ is infinite-dimensional,  and  can  be  naturally
identified  with  the algebra of polynomial functions on the
quantum 2-sphere $S_{\mu}^2$ [P].

    (B) Classical principal bundles provide a natural mathematical
framework for the study of gauge theories. It is interesting to see what
will be the counterparts of these theories, in the context of quantum
principal bundles [D3] ($M$ playing the role of the space-time).
Properties of such 'quantum gauge' theories
essentially
depend (besides on the 'symmetry group' $G$), on the following two
prespecifications:

As first, it is necessary to fix a (bicovariant *-) calculus
$(\Gamma,d)$ over $G$. This determines kinematical degrees of freedom.
Secondly, we have to choose a map
$\delta:\Gamma_{inv}\ra\Gamma_{inv}^{\otimes 2}$. This influences dynamical
properties of the theory, because $\delta$ implicitely figures in the
expression for the curvature.

Closely related with problematics of quantum gauge theories
is the question of 'gauge transformations'. If $M$ is a
classical smooth manifold then the most
direct way of defining gauge transformations as automorphisms of the bundle
$P$ gives nothing new, because of the inherent geometrical inhomogenity
of the bundle $P$. More precisely, automorphism groups of $P$ and
its classical part $P_{cl}$ are isomorphic. However, a proper
quantum generalization of gauge transformations can be introduced via
the concepts of quantum (infinitezimal) gauge bundles [D2,3]. These are bundles
associated to $P$, relative to the adjoint actions of $G$ on $G$ and
$\Gamma_{inv}$ respectively.

    (C) Interesting examples of quantum principal bundles  can
be obtained from quantum homogeneous spaces. A general construction
is this [D2]. Let $G'$ be a compact matrix quantum  group.  Entities
related to $G'$ will be endowed with a prime. Let us assume  that  $G $
is a subgroup  of $G'$.  At the formal  level,  this  presumes  a
specification of a *-epimorphism $q:{\cal A}'\ra {\cal A}$
such that
\[(q\otimes q){\phi}'={\phi}q\]
\[kq=qk'.\]

     The *-homomorphism $F:{\cal A}'\ra {\cal A}'\otimes {\cal A}$
given by
                          \[F=(id\otimes q)\phi'\]
is interpretable as the right action of $G$ on  $G'$.  Let $M$  be  the
corresponding 'orbit space'. This  space  is  represented  by  the
fixed point *-subalgebra ${\cal V}$.
Let $i:{\cal V}\hookrightarrow {\cal A}'$ be the  inclusion map.
The triplet $P=({\cal A}',i,F)$ is a quantum principal $G$-bundle over $M$.
Because of $\phi'({\cal V})\subseteq {\cal A}'\otimes {\cal V}$
there exists a natural left action of $G'$ on $M$,
represented by $\phi'i:{\cal V}\ra {\cal A}'\otimes {\cal V}$
($M$ is a quantum homogeneous $G'$-space).
\filbreak
    Let $(\Gamma',d')$ be a bicovariant first-order *-calculus
over $ G'$, and ${\cal R}'\subseteq ker(e')$ the corresponding right
${\cal A}'$-ideal.  Let  us
assume that
\[q({\cal R}')\subseteq{\cal R},\]
where ${\cal R}\subseteq ker(e)$ is the right ${\cal A}$-ideal
determining the  calculus  $(\Gamma,d)$
over $G$. Then the map $q:{\cal A}'\ra {\cal A}$ can be
(uniquely) extended to  the  (*-) homomorphism
${\hat q}:\Gamma'^\wedge\ra\Gamma^\wedge$  of differential algebras.

Let $\Omega(P)$ be an arbitrary graded differential *-algebra built
over $(\Gamma',d')$, satisfying (ii) of Section II.
The  differential algebra $\Gamma'^\wedge$
possesses this property.  In this particular case
${\hat F}=(id\otimes {\hat q}){\hat \phi}'.$

     Let us assume that a linear map
$\epsilon:\Gamma_{inv}\ra\Gamma_{inv}'$ is given, such that the following
identities are satisfied:
\[(id\otimes q)\widetilde{ad}'\epsilon =(\epsilon\otimes id)\widetilde{ad}\]
\[{\hat q}\epsilon (\vartheta)=\vartheta.\]
Then
the map $\omega:\Gamma_{inv}\ra\Omega^1 (P)$,  obtained by
composing $\epsilon$ with
the  cannonical inclusion $\Gamma'_{inv}\hookrightarrow\Omega^1(P)$,
is a connection on $P$. Further, if
\[\epsilon(\vartheta\circ q(a))=\epsilon(\vartheta)\circ a\]
then $\epsilon\otimes\epsilon$ intertwines cannonical flip-over operators.
Finally, if in addition
\[d'\epsilon(\vartheta)=0\]
then the commutation property
\[\omega (\vartheta)\varphi=
(-1)^{deg(\varphi)}\sum_{k}\varphi_k\omega(\vartheta\circ c_k)\]
holds for each $\varphi\in \Omega(P)$. In particular, $\omega$ is regular.

As a concrete illustration, let us briefly consider the  case
$G'=SU_\mu (2)$ $(\mu\in (-1,1)\setminus\{0\})$ and
$G=G'_{cl} =U(1)$. This  gives  the  quantum
Hopf fibering $SU_\mu (2)\ra S^2_\mu$ .
Let us assume that the calculus on  $G$  is the standard one,
based on differential forms. A bicovariant calculus
$(\Gamma',d')$ over $G'$ satisfies $q({\cal R}')\subseteq{\cal R}$
iff it is admissible, in  the  sense mentioned in (A).
Let us assume that $(\Gamma',d')$ is a
minimal admissible calculus, and let $\Gamma'_{inv}$ be identified with  the
polynomial algebra over $S_\mu^2$. Let $\tau\in{\Gamma'}_{inv}$ be the
element corresponding to the unit function on $S_\mu^2$ and let us define
\[\Omega(P)=\Gamma'^\wedge/\{d'\tau=0\}.\]
The space $\Gamma_{inv}$ is
1-dimensional and generated by ${\hat q}(\tau)$. A map $\epsilon$
can be defined by $\epsilon{\hat q}(\tau)=\tau$. It is worth noticing that
the curvature of the corresponding connection $\omega$ vanishes.

     REFERENCES
     \begin{itemize}
     \item[(C1)] Connes, A. {\em Non-commutative   differential    geometry.}
Institut des Hautes Etudes Scientifiques: Extrait des Publications
Mathématiques No.62 (1986);
     \item[(C2)] Connes, A. {\em Geometrie  non  commutative.}  InterEditions,
Paris (1990);
    \item[(D1)] \Dj ur\dj evi\'c , M. {\em Geometry  of  Quantum  Principal
Bundles  I.}
Preprint QmmP 6/92, Belgrade University.
    \item[(D2)] \Dj ur\dj evi\'c, M. {\em Geometry of Quantum Principal Bundles
II.}
Preprint QmmP 4/93, Belgrade University.
\item[(D3)] \Dj ur\dj evi\'c , M. {\em Quantum principal bundles and
corresponding
gauge theories}. Preprint QmmP 2/93, Belgrade University.
   \item[(KN)] Kobayashi, S. and Nomizu,  K. {\em Foundations of Differential
geometry}. Interscience Publishers, New York London (1963);
    \item[(P)] Podles, P. {\em Quantum Spheres}. Lett. Math. Phys.  14  193-202
(1987);
   \item[(W1)] Woronowicz, S.L. {\em Twisted SU(2) group. An example of a  non-
commutative differential  calculus.}  RIMS,  Kyoto  University  23,
117-181, (1987);
   \item[(W2)] Woronowicz, S.L. {\em Compact  Matrix  Pseudogroups.}  CMP  111,
613-665, (1987);
   \item[(W3)] Woronowicz, S.L. {\em Differential Calculus  on  Compact  Matrix
Pseudogroups (Quantum Groups)}. CMP 122, 125-170, (1989);
\end{itemize}
\filbreak

     ACKNOWLEDGMENTS

     This paper is an extended form of my lecture at the XXII-th
Conference on
Differential Geometric Methods in Theoretical Physics, Ixtapa-Zihuatanejo,
Mexico, Septembre 1993.

     The possibility of my coming in Kingdom of the Feathered Serpent, and of
my attendance on the Conference, is based on the united financial
support of many firms and citizens from my hometown Petrovac-na-Mlavi, in
Serbia:

\begin{tabbing}
mmmmmmmm\=\kill
\>--Private company "Vlaji\a'c-Komerc": Tomi\v{s}a Vlaji\a'c\\
\>--Mrs Milica Andrejevi\a'c, and Miss Tijana Andrejevi\a'c\\
\>--Dentist Ordination of Dr Radivoje Karad\v{z}i\a'c, and
Mrs Sne\v{z}ana Karad\v{z}i\a'c\\
\>--Social company for remaking plastic materials "Mlavaplastika"\\
\>--Social tannery "Ikop"\\
\>--Social furniture factory "Javor"\\
\>--Agricultural corporation "Borac"\\
\>--Private savings-bank "Sinkom": Stojkovi\a'c Sini\v{s}a\\
\>--Mrs Filipovi\a'c Desanka;--Mr Ugrinovi\a'c Miroljub\\
\>--Private firm "Zlatar": Dini\a'c Du\v{s}ko\\
\>--Apothecary's shop "Apoteka": \Dj or\dj evi\a'c Goran\\
\>--Mr Stoimirovi\a'c \v{Z}ika; --Mr \Dj or\dj evi\a'c Predrag\\
\>--Private firm "\v{Z}arac": \Dj or\dj evi\a'c Branko   \\
\>--Private firm "MSM": Ran\v{c}i\a'c Slavi\v{s}a          \\
\>--Apothecary's shop "Apoteka \v{S}eki": Milo\v{s}evi\a'c Dragi\v{s}a\\
\>--Private firm "Lord": Aleksi\a'c Tale\\
\>--Hairdresser saloon "Frizer": Savi\a'c Peri\v{s}a-Slav\v{c}e\\
\>--Coffee shop "Man": Milovanovi\a'c Neboj\v{s}a\\
\>--Cabinet-maker's workshop: Kosti\a'c Jovan\\
\>--Building material Store: Milovanovi\a'c Mi\v{s}a\\
\>--Furniture shop "Hrast"\\
\>--Cake shop "Havaji": Pani\a'c Radomir\\
\>--Mr Milosavljevi\a'c Vlastimir; --Mr Mi\v{c}i\a'c Dragan\\
\end{tabbing}

Warm thanks are due to all of them.

     I am especially grateful to Mrs Milica Andrejevi\'c,  Mr Radivoje
Karad\v{z}i\'c and Mrs Snezana Karad\v{z}i\'c, and last but not least to Mr
Obren Joksimovi\'c, for their care and
interest about  my work, continuous support, and organization of
this sponsorship.

I would like to thank to the Soros Yugoslavia Foundation, and to the
Organizers of the XXII-DGM Conference for partial financial supports.
\end{document}